# Cosmological Principle and Relativity - Part I

## A G Oliveira[1] and R Abreu[2,3]

[1]Direcção Técnica, Radiodifusão Portuguesa, Lisboa, Portugal
[2]Departamento de Física, IST, Lisboa, Portugal
[3]Centro de Electrodinâmica, IST, Lisboa, Portugal
E-mail: alfredooliveira@rdp.pt and rabreu@ist.utl.pt

**Abstract.** The absence of an identified consequence at solar system scale of the cosmological space expansion is usually explained considering that space expansion does not affect local anysotropies in matter distribution. This can also be explained considering a time dependent scenario compatible with Relativity Principle, therefore supporting physical laws independent of the time position of the observer. A theory considering such relativist scenario, i.e., generalizing Relativity Principle to position, embodies Cosmological Principle and can be intrinsically able to fit directly both local and cosmic data. In part I it is presented the general framework of such a theory, called Local Relativity (LR), and analysed the space-time structure. Special Relativity space-time is obtained, with no formal conflict with Einstein analysis, but fully solving apparent paradoxes and conceptual difficulties, including the simultaneity concept and the long discussed Sagnac effect. In part II, LR is applied to positional analysis. It is verified the accordance with solar system measurements and with classic cosmic tests, without dark matter or dark energy. Two of the new features obtained in part II are the possibility of a planetary orbital evolution compatible with a null determination for $G$ variation, supporting a warmer scenario for earth (and Mars) past climate, and the possibility of an accelerating component in earth rotation, compatible with the most recent measurements.

## 1. Introduction

Most cosmological models consider a variation with time of matter/space characteristics, either considering the time variation of some fundamental physical parameter (i.e., a fundamental magnitude or constant) or/and a cosmological line element affected by a time dependent scale factor, as the standard model for cosmology. A problem of such a scenario is that the validity of fundamental physical laws and theories through time is not clear because they ignore the possibility and the consequences of a variation of matter/space with time. Einstein [1] considered that Relativity Theory is not valid in a space expansion scenario because this one conflicts with fundamental aspects of Special Relativity, namely the non-local determination of the time coordinate. Note that Einstein Relativity does not impose the time invariance of matter/space characteristics, namely the time constancy of light speed. Such possibility was simply not considered. At the time, there were no observational results able of supporting it. As it is evident, there is no incompatibility between Relativity Principle and a positional (Time/Space) variation of matter/space able of supporting physical laws independent of the position of the observer. Relativity Principle states that physical laws, or "laws of nature", can be represented as a function only of measures relative to the observer. Einstein determined its consequences, in what concerns motion and gravitational field but not in what concerns position in Time or Space. This situation is similar to the relation between Special and General Relativity: Special Relativity did not considered gravitational fields but that does not mean that Special Relativity implies that a motion is not affected by the presence of a gravitational field. What it means is that Special Relativity is valid "only so long as we are able to disregard the influences of gravitational fields on phenomena (e.g. of light)." [2]

To build a cosmological theory, something has to be stated on how phenomena depend on position. Cosmological Principle states that there is nothing special about this particular point of the Universe where we are. This is an important anti-anthropomorphic statement. However, it is only that. It does not imply that the Universe has to be uniform, or isotropic, or that phenomena have to follow the same laws whatever the point of the Universe. Statements like these, as well as the presumption of the time variation of some physical parameter, are just hypothesis, not supported in experience. Hence, we conclude that the fundamental assumptions of present cosmological theories are mainly a consequence of the absence of information and have no direct support from experience. This is not a problem, it is just a methodology for facing the situation; the expectation is to obtain a theory liable to fit observations. However, even the current standard model needs additional hypothesis, namely dark matter, dark energy, inflation phase, which have also no direct support from experience.

Efforts have been made to measure at solar system scale consequences of the presumed space expansion or a time variation on the value of fundamental constants. The result is negative, namely in the analysis of the range data to the Viking landers on Mars [3, 4]. This is an important result because it indicates that if there is a time dependent phenomenon, either a space expansion or other variation of matter/space, then it is probably able to support physical laws independent of position, i.e., such eventual variation probably obeys Relativity Principle. For the first time, we can state something about the dependence of phenomena with time supported on experience. This enables us to consider that phenomena can be described by laws dependent only on measures relative to the observer, whatever his motion, field or position on space and time. *This generalization of Relativity Principle embodies Cosmological Principle.*

The above generalization of Relativity Principle implies the possibility of a variation of matter/space characteristics with position, as it happens with motion and field. In such case, it is advisable to consider that the data supporting physical laws can only be classified as local because the distances and time intervals involved are too small for detecting an eventual variation of matter/space characteristics with space or time position. Therefore, we consider, this being a key aspect of our analysis, that current fundamental physical laws are locally valid but we do not presume, «a priori», they have non-local validity. In accordance with this line of thought, we consider also that there is only experimental support for considering that Relativity Principle applies to local laws. This does not mean that it does not apply non-locally either, only that one cannot presume it «a priori». To the theory this way obtained we called Local Relativity.

In this work we first establish the general framework of Local Relativity. This framework defines how matter/space properties can vary but not how do they actually vary. To find that out, one has to analyse the consequences of motion, field and position of the observer. Such analysis imply the interpretation of the results of non-local observations, therefore requires a clear understanding of space-time. The space-time structure is also established in this work. The positional analysis is necessary to test our generalization of relativity property, being therefore also presented. The motion and field analyses, as well as the development of cosmological models, is not the object of present work

This work is divided in two parts. In this part I, we define measuring units and observers, characterise the relativity property and determine the relations between physical parameters imposed by it. The properties of light velocity are carefully characterized. They are all that is needed to obtain a trivial explanation for Sagnac effect and to obtain the special relativity space-time continuum, now with a clear understanding of the physical relation between Time and Space. This part ends with the analysis of reference frames, indispensable for non-local analysis. Time and Length paradoxes of special relativity are then fully solved. An important achievement of this part I is that we obtain the two principles of Special Relativity as a consequence of the constancy of mean light speed in a closed path, considering a 3-Euclidian space with no connection between time and space. This leads to a new understanding of the Universe.

In part II, a time varying scenario for matter/space is defined after redshift and cosmic microwave background characteristics. The direct consequences, without any additional hypothesis like dark matter or dark energy, at cosmic and solar system levels, are presented. It is verified that they are not contrary to any known observational data and new, testable, results are obtained.

## 2. Local Relativity

The analysis we present next is not an adaptation of Einstein work. Local Relativity is deducted independently, exclusively based on relevant experimental results and not on any of the results of Einstein Relativity or in the concepts of Space and Time developed after them. The reader is asked the effort of putting aside all that by now, in order to be able to properly follow our line of thought. Furthermore, what we present here is not a theory developed from hypotheses but, on the contrary, a research carried on experimental results, sticking as close as possible to their range of validity and avoiding any statement or hypothesis not conveniently supported in them. To the characteristic behaviours of phenomena we call "properties" and do not classify them as "principles" because they are the consequence of something, even if we do not know what. When this research is finished, the findings can then be framed in a theory deducted from certain fundamental assumptions.

### 2.1. Local Laws and Local Relativity Principle.

An essential aspect is that we do not presume «a priori» the non-local validity of current physical laws, considering that we can only attribute local validity to the data on which they are based. In what concerns current fundamental physical laws, we consider that they are based only on local observations and can be considered as **local physical laws**, i.e., *the limit of* *general physical laws when r→0 and dt→0 in the observer's reference frame.* Note that this analysis is not intended to be valid at sub-atomic level.

In what concerns the relativity property displayed by phenomena, what can be confidently stated is that local physical laws seem to be valid whatever the inertial motion of the observer in our time and space neighbourhood. Note that the range of velocities of the observer is very small in relation to light speed and one cannot state that laws of mechanics do not neglect an eventual influence of observer's motion. In order to obtain a theory for any point of space and time, it is necessary to generalise the above statement. As the successive local observations, although relative to a very small range in space and time, have shown no change in the results, what can be stated, considering the above, is that:

- *Local physical laws can represent validly, to the first approximation, the instantaneous relations between the local measures made by an atomic observer in a local physical system, whatever the inertial motion, gravitational field and position in time and space of the atomic observer.*

We name this enunciation of Relativity Principle in our local framework as **Local Relativity Property** (LRP).

## 2.2. Measuring units

In order to analyse situations where physical parameters may vary, a clear definition of measuring units is indispensable. We will use as fundamental magnitudes Mass, Length, Charge and Time. We will consider that Mass, Length and Charge are measured by comparison with bodies chosen as standard. This does not have to correspond to the practical realisation of measurements; however, whatever the method used, the results must correspond to the ones obtained this way. The magnitude Time is intrinsically different from the other magnitudes. The crucial property of Time unit is that *Time unit must satisfy the invariance of local physical laws*, otherwise there will be a violation of Local Relativity Property. As will be shown later, such invariance is satisfied provided that the measures of fundamental constants keep invariant. Therefore, one can define time unit from the constancy of the measure of a convenient physical constant.

## 2.3. Constancy of Local Mean Light Speed and Time Unit

In Maxwell electromagnetic theory, light speed in vacuum has the value of the electrodynamic constant $c_0$. As this constant has the dimensions of a velocity, it seems to be quite suitable for defining time unit. However, one must note that until now only the mean value of light speed in a closed circuit has been directly measured, this being the experimental result in which one can support the definition of time unit.

In Local Relativity framework, in order to interpret such experimental result, we define **local mean light speed** as *the limit of the measure of mean light speed in a closed path when the path length tends to zero.*

Michelson-Morley [5] experiment showed that local mean light speed is independent of the direction of light ray for an atomic observer at rest on earth surface; Kennedy-Thorndike [6] experiment showed that it is also independent of the inertial motion. More recent experiments, using lasers and cavities (note that these are standing-wave devices, therefore with a characteristic period dependent on a closed wave path), namely Brillet and Hall [7], have also concluded that *the local mean light speed measured by an atomic observer is* *constant, whatever the inertial motion of the observer.* This conclusion is not entirely consensual [8]. Nevertheless, we will consider the above statement a local physical law, valid to the first approximation in our time and space neighbourhood. We will designate it by **Michelson Law**.

One can note that Michelson law has a trivial explanation if light speed depends on light source motion ("ballistic" hypothesis) but not if light speed is independent of source motion.

Local Relativity Property requires Michelson law to be valid whatever the inertial motion, field or position of the observer; therefore, one can define time unit stating that **Time Unit** *is such that the local mean light speed, measured by an atomic observer, is* $c_0$.

## 2.4. Velocity – the concept and the measure

The concepts of speed or velocity are linked to the ratio between a path length and the time to cover it, measured with observer's length and time units. This is enough to make measures of the mean speed of a radiation or of a body performing a cyclic motion in the neighbourhood of the observer. The observer measures the path using a measuring rod at rest in relation to him and measures with his local clock the time interval by noting the time instants he acknowledge two successive passages of the body or radiation in the same point of space in relation to him. However, to measure a speed or a velocity between two non-coincident points of space, it is necessary to relate time in the two points with observer's local time, i.e., to attribute time coordinates to those points or to synchronise clocks placed at them. Let us remind the fundamental aspects of the problem.

For measuring the relative velocity of a body much slower than light, one can neglect the time light spends between two measuring points, attributing to the passage of the body in each of the points a time coordinate which is the observer's local time when he sees the occurrences. Or, if one wants to be more precise, one can attribute to the one-way light speed the value $c_0$ of the mean light speed in a closed path, presuming that the one-way value will not be very different from this mean value. However, the faster the body (in relation to the observer), the greater the influence of the value one attributes to light velocity in the value obtained for the velocity of the body. At the limit, if one intends to measure the one-way light speed, the value obtained is independent of the true one-way light speed and depends only on the mean value for light speed in closed path, a result we detail in the next paragraph. This implies that an observer cannot make a "direct" measure of one-way light speed, unless he can use something much faster than light.

In his article of 1905 [9], Einstein presented a method for clocks synchronisation. The solution presented by Einstein is to establish *as definition* that the time light spends from a point $A$ to a point $B$ is equal to the time light spends from $B$ to $A$ and to assume that the mean value (path $A{\rightarrow}B{\rightarrow}A$) of light speed is $c_0$. We will now show that this Einstein symmetric synchronisation is possible whatever the values of light speed in each of paths $A{\rightarrow}B$ and $B{\rightarrow}A$, provided that its mean value keeps constant. To analyse the process with generality, consider that light speed is $v_1$ from $A$ to $B$ and $v_2$ from $B$ to $A$. The clock at $A$ is identical to the clock at $B$ but marks a different time, the difference being $\Delta t$. The observers at $B$ and $A$ see the clocks marking a difference of, respectively:

$$t_B - t_A = -\Delta t + \frac{\overline{AB}}{v_1} \qquad t'_A - t_B = \Delta t + \frac{\overline{AB}}{v_2} \tag{1}$$

where $\overline{AB}$ is the distance between $A$ and $B$. The clocks are considered synchronised when $t_B - t_A = t'_A - t_B$, therefore:

$$\Delta t = \frac{\overline{AB}}{2}\left(\frac{1}{v_1} - \frac{1}{v_2}\right) \tag{2}$$

Then, the observers see a difference between clocks of:

$$t_B - t_A = t'_A - t_B = \frac{\overline{AB}}{2}\left(\frac{1}{v_1} + \frac{1}{v_2}\right) = \frac{\overline{AB}}{\overline{v}} \tag{3}$$

where $\overline{v} = 2\left(v_1^{-1} + v_2^{-1}\right)^{-1} = 2\overline{AB}\left(t'_A - t_A\right)^{-1}$ is the mean light speed in the path $A{\rightarrow}B{\rightarrow}A$. Therefore, when an observer at $A$ or $B$ measures light speed after performing a symmetric synchronisation of clocks, *what is obtained is the mean value of light speed in the closed path $A{\rightarrow}B{\rightarrow}A$ and not its one-way value.* One can now note that the validity of (3) in relation to a third clock requires the constancy of the mean light speed $\overline{v}$. Therefore, Einstein synchronisation is insensitive to the one-way light speed provided that its mean value keeps invariant. After having defined the synchronism, Einstein considered, in agreement with experience, that this mean light speed in a closed path is a universal constant – the velocity of light in empty space.

Given that the one-way light speed is not known, one could think of other simple methods of attributing time coordinates or synchronising clocks. Such a method could be to take two identical clocks and then to move them symmetrically from the observer to two points at the same distance of him. The problem is that the fact they are moved symmetrically in relation to the observer does not guarantee, «a priori», that they are equally affected during the motion. Therefore, one can imagine different simple methods of determining time coordinates, leading to different values for the time coordinate of a point, but one cannot establish «a priori» a criterion for validation.

One can also make indirect determinations of a speed, for instance, measuring a wavelength shift. However, the relation between the wavelength shift and speed depends on some physical theory, and different relations can arise from different theories. The same reasoning applies to methods derived from cosmological observations.

Finally, one can search for more ingenious methods of synchronizing clocks, like Sama's [10]. The fact is that a practical and consensual method has not yet been put through.

Considering the above, we conclude that we can establish a concept for speed or velocity as the ratio between a path length and the time to cover it in observer's units and that this concept can be implemented in mean speed measures in closed paths; in what concerns opens paths, one has not, until now, established any definition of velocity with a known correspondence with the concept above defined. That does not prevent us from measuring velocities by determining the ratio of a variation of length and time coordinates. However the value so obtained depends on the method chosen for determining time coordinates and, therefore, we have to identify it when referring to velocity measures. In this paper we will consider Atomic velocities, Newton velocities and Einstein velocities, to be defined later. Naturally, *the measures of speed in closed paths are the same whatever the method but the one-way measures of velocity can be different.*

## 2.5. Two observers

For the interpretation of observations, the correct characterization of observer properties is of crucial importance. The same happens when establishing a physical theory. In this work we will use two different observers.

Local observations can be described as being snapshots of the relative appearance of local phenomena to a local atomic observer. Therefore, for the analysis of local observations and of local physical laws it is convenient to consider an atomic observer, **the observer A**, such that *a local atom is invariant in relation to him*, whatever his motion, field and position. As he is intended to represent our instruments and us, the *A* observer is dependent on light speed for acknowledging an occurrence.

In cosmic observations, the same observer watches phenomena in different points in time and space. Therefore, cosmic observations may reflect the dependence with position (time/space) of physical laws in relation to an *invariant* observer. Therefore, we will also consider a non-atomic observer, **the observer R**, such that *the relations between the measuring units of two R observers are invariant*, whatever their relative motion, field or position.

The problem of attributing time coordinates, as we have already noted, lies in the fact that an atomic observer depends on the unknown light velocity to acknowledge an occurrence. A way of dealing theoretically with this problem is to introduce an observer that depends on something much faster than light. At the limit, a solution is to introduce an observer with "instant vision". To achieve that, we will consider that *R observer has "instant vision"*, i.e., he is not dependent on light velocity to acknowledge an occurrence.

One must note that observers are conceptual entities. One can emulate an *A* observer but not an *R* observer. That is irrelevant for the analysis.

## 2.6. Notation

As we will use two different observers, measures and measuring units have to be carefully identified. The units of measure of fundamental magnitudes are usually represented by the same letter that represents the magnitude and the units of other physical parameters by the symbol of the parameter between brackets. This convention is followed in the paper, with the addition of a superscript, identifying the observer to whom the unit belongs. The subscript identifies the observer that measured the parameter. So, the fundamental units of measure of *A* are $M^A$, $Q^A$, $L^A$ and $T^A$; $[G]^A$ represents the *A* measuring unit of *G*; $G_A$ is the value of *G* measured by *A*; $M^A_R$ represents the value of the mass unit of *A* measured by *R* ($M^A_R = M^A/M^R$); $[G]^A_R = [G]^A / [G]^R$ is the value of the *A* unit of *G* measured by *R*.

The measuring units of *R* and *A* are equal at the point in time and space chosen as origin, named "zero point", considering an absence of field at the point and that both observers are "at absolute rest". The precise meaning of this concept is established later. The local measures made in this situation are identical whatever the observer, being identified by the subscript "0". LRP implies that the *A* local value of physical parameters is always the same as the value measured in this situation, for instance, $G_A = G_0$.

## 2.7. Local Relativity Conditions: physical parameters interdependence

LRP states the invariance of our perception of local phenomena. This does not imply the invariance of physical parameters, only of our measure of them. In order to understand what this invariance implies, let us analyse the measuring procedure.

The variations of Mass, Length and Charge, if the number of atomic particles do not change, cannot be directly detected by an atomic observer because they are measured comparing a body with a reference body; a variation of any of those parameters affects both bodies but not the result of the comparison. In what concerns Time, as Time unit is chosen by the atomic observer so that physical laws keep invariant, it cannot vary if everything else keeps invariant. So, what happens if a fundamental magnitude changes? An atomic observer will conclude that a fundamental constant has changed. For instance, if Mass doubles, the perception of an atomic observer will be that gravitational constant $G$ has changed because he will measure a different acceleration between the same bodies at the same distance. However, what changed was not $G$ but only his measure of $G$, because the measuring unit $[G]$ changed. From Newton laws, this measuring unit is related with the units of Mass, Length and Time by the dimensional equation

$$[G] = M^{-1}L^3T^{-2} \tag{4}$$

The equation shows that if Mass unit doubles and no other magnitude or constant changes, the unit of measure of $G$ becomes 1/2; therefore, the value determined for $G$ doubles, provided that Newton laws keep valid. Now, consider that Mass doubles and $G$ reduces to 1/2. Then, the value determined for $G$ is the same: an atomic observer measures the same acceleration between the same bodies at the same distance. Once an atomic observer cannot directly measure a variation in fundamental magnitudes, if his local measurements of fundamental constants do not display any variation then his description of local phenomena keeps unchanged. We conclude that *the constancy of local measurements of fundamental constants by an atomic observer* is enough condition for LRP.

To determine the relations between fundamental physical parameters that ensure their variations do not produce detectable consequences on local phenomena is, therefore, to establish the conditions that ensure the invariance of the local measure of fundamental constants, namely $G$, $\varepsilon$ and the local mean light speed $c$. The obvious solution is that fundamental constants can only vary accordingly with their dimensional equations. We will detail the deduction of this result. Their local measures made by an atomic observer $A$ are:

$$G_A = G_0 \qquad \varepsilon_A = \varepsilon_0 \qquad c_A = c_0 \tag{5}$$

The values of fundamental constants to an $R$ observer coincident with the $A$ observer have not to be the same because $A$ measuring units depend on motion, field and position. As measures are inversely proportional to measuring units, representing a generic physical entity by *phy*, it is:

$$phy_R = phy_A \cdot [phy]_R^A \tag{6}$$

The relations between $R$ and $A$ measures of fundamental constants are then:

$$\frac{G_R}{G_A} = \left(M_R^A\right)^{-1}\left(L_R^A\right)^3\left(T_R^A\right)^{-2}$$

$$\frac{\varepsilon_R}{\varepsilon_A} = \left(Q_R^A\right)^2\left(M_R^A\right)^{-1}\left(L_R^A\right)^{-3}\left(T_R^A\right)^2 \qquad (7)$$

$$\frac{c_R}{c_A} = L_R^A\left(T_R^A\right)^{-1}$$

The objective is to define how fundamental magnitudes and constants can vary in relation to their values in the "zero point". We will represent this relative variation by the identifying letter of the physical parameter without scripts. Its value is the ratio between $A$ and $R$ measuring units, for instance $M \equiv M_R^A$, or the ratio between $R$ and $A$ measures, for instance, $c \equiv c_R/c_A = c_R/c_0$. Rearranging equations (7) one obtains:

$$T = c^{-1}L$$

$$GM = c^2L \qquad (8)$$

$$\sqrt{G/\varepsilon}\,Q = c^2L$$

Conditions (8) are called LR conditions. They are the relations between fundamental physical parameters that ensure the invariance of their local measure by an atomic observer.

An important consequence of (8) is that, in order to obey LRP, either all physical parameters are invariant or at least two of them have to vary. Cosmological theories that consider a variation of only one parameter imply a variation in local phenomena, which is not supported by observations. On the contrary, any matter/space varying scenario to which corresponds a variation of fundamental physical parameters in accordance with LR conditions obeys Relativity Principle at local scale, therefore being not locally detectable by an atomic observer but being able to produce observable consequences at non-local scale.

### 2.8. Variation of Magnitudes and Constants (in R)

As LRP implies the invariance of the $A$ measure of any physical entity, i.e., it implies $phy_A = constant$, then, from (6), the variation of $phy_R$ is equal to the variation of the $A$ measuring units of $phy$ in $R$, $\left[phy\right]_R^A$. In other words, physical entities have to vary accordingly with their dimensional equations, as already noted. For instance, the wavelength (in $R$) of a spectral radiation has to vary accordingly to $L$ because $\left[\lambda\right]_R^A = L$. One may wonder what can be the relation between both but there is an easy answer: they are both related, directly or indirectly, with the size of electronic orbitals. Also, the energy of emitted radiation in $R$ has to vary with energy dimension $ML^2T^{-2} = Mc^2$, the Planck constant with $McL$. Only in this way can phenomena keep invariant to $A$, as stated by the LRP.

When an $A$ observer makes non-local observations, the variation (in $R$) of non-local phenomena and of $A$ measuring units are no longer equivalent, therefore $A$ can detect a relative difference. For instance, in case of a $L$ variation with time, $A$ will detect a wavelength difference between a radiation from a distant source and the correspondent local radiation, this considering that there is no change in the wavelength from the source to $A$. On the other hand, dimensionless numbers, like the fine structure constant, have naturally to keep invariant in $R$ and, therefore, in $A$ non-local observations (considering that there is no change in light properties in the path between source and observer). The *invariance of dimensionless local combinations of physical parameters* can also be used as an enunciation of LRP. Albrecht and Magueijo [11] have already referred that physical experiments are only sensitive to dimensionless combinations of dimensional physical parameters. The non-local analysis of such numbers, namely the analysis of the fine structure constant at cosmic distances [12, 13, 14, 15], in different gravitational fields or at different velocities, are positional, field and dynamic tests for LRP and source of information on the change of light properties between the source and the observer.

### 2.9. Non-local analysis

To perform a non-local analysis implies the capability of attributing a time and a space coordinate to each point, i.e., to define a reference frame and to establish the correspondence between the observation of an occurrence and its position in the reference frame. Next chapters are concerned with this problem.

## 3. Time and Space

As already noted, the constancy of local mean light speed is a trivial result if light speed directly depends on source motion. However, if this is not the case, then, in LR framework, this law implies that the velocity unit of an atomic observer has to vary with the motion of the observer, i.e., length and/or time atomic units cannot be invariant. In this case, time and space for an atomic observer become dependent on his motion. This reasoning applies also to field and position, which cases will not be analysed here. Therefore, the first step is to establish how light speed depends on source motion.

### 3.1 Light speed and the state of motion of the emitting body

An important number of experimental and observational results, starting with 1910 Comstock considerations on the orbits of close binary stars and Tolman experiment [16, 17], being specially relevant W. de Sitter [18] analysis of binary systems of spectroscopic twin stars, have systematically concluded that light speed is independent of the motion of its source, although objections to some results have been presented. A report can be found in Zhang [19] and in [20]. Note that, according to Zhang, «the "independence" does not mean that either the one-way velocity $c_r$ of light is equal to the two-way velocity $c$ of light or the one-way velocity of light is isotropic». At this point of our analysis, we are not yet in conditions to interpret non-local results, so we need a local one. Such a local result can be the Sagnac effect [21, 22, 23]. A simple way of accounting for this effect is to consider that light has a one-way speed of $c_0$ in the lab reference frame, whatever the speed of the source of light. However, if light speed is independent of source motion, why should it be constant in relation to the lab? One can state that it is so, supported in the absence of contrary evidence, but one must certainly look for other explanation of Sagnac effect. Special Relativity cannot be used at this point because our aim is to explain it from the fundamental results obtained so far.

As already referred, what can be stated about the value of light speed is that its local mean value measured by an atomic observer is $c_0$, therefore, that the mean light speed in a closed (and small) circuit is $c_0$ in the lab $A$ reference frame (we call this the Michelson law). Analysing Sagnac effect, one can note that the displacement of the interferometer produces a fringe phase shift that depends on the time spent by light in covering the displacement path in both ways, i.e., in a closed path. This conclusion can be illustrated analysing a ring interferometer of radius $a$ rotating with angular velocity $\Omega$ (figure 1). The one-way light speed at opposite ways from the beam splitter/combiner are $v_1$ and $v_2$ in the $A$ lab reference frame. Then, the time $\Delta t_1$ and $\Delta t_2$ spent by light rays circulating the ring in one and in the other way is the time for circling the interferometer plus or minus the time for covering the displacement path. As the first one is a closed path, the mean light speed in it is $c_0$ and:

$$\Delta t_1 = \frac{2\pi a}{c_0} + \frac{\Omega a \cdot \Delta t_1}{v_1} \qquad \Delta t_2 = \frac{2\pi a}{c_0} - \frac{\Omega a \cdot \Delta t_2}{v_2} \qquad (9)$$

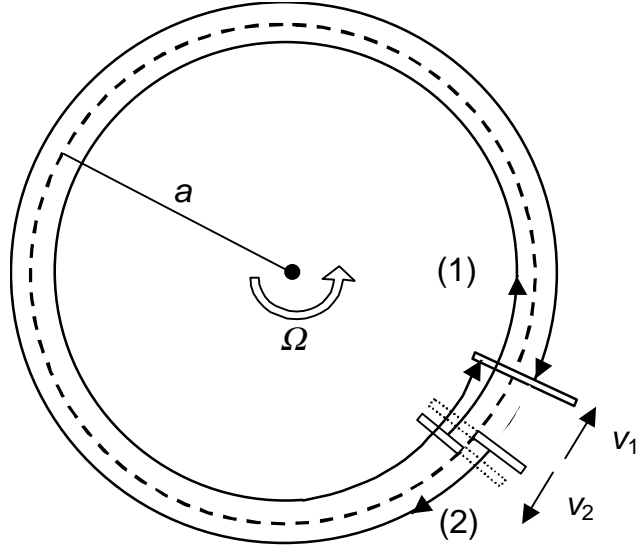

**Figure 1**. Model of the Sagnac interferometer. The figure represents the paths of two light rays that reach the beam splitter/combiner at the same instant from opposite directions, in the lab $A$ reference frame. The positions of the beam splitter when each of light rays was originated are shown with half dotted. Light speed in the lab $A$ reference frame is $v_1$ and $v_2$ at opposite directions from the beam splitter. Total path of each of light rays (1) or (2) is the circle of radius $a$ plus or minus the displacement of the beam splitter. The first is a closed path, where the mean light speed is $c_0$ in the $A$ lab reference frame, therefore with the same time length for both light rays. The time difference between the two light rays is then, approximately, the time spent by light to cover the displacement of the beam splitter/combiner in opposite ways, therefore with a mean light speed of $c_0$, whatever $v_1$ and $v_2$. Sagnac interferometer does not give any indication on the one-way light speed but it shows that light speed is independent of source motion.

The first order time difference is:

$$\delta t = \Delta t_1 - \Delta t_2 \approx \frac{4A\Omega}{c_0 \overline{v}} \Leftrightarrow \frac{4A\Omega}{c_0^{\,2}} \qquad (10)$$

where $A$ is the area of the interferometer and $\bar{v} = 2.(v_1^{-1} + v_2^{-1})^{-1}$ is the mean value of the one and the other way light speed in the displacement path; hence, first order results of Sagnac effect imply $\bar{v} = c_0$ in the $A$ lab reference frame, whatever $\Omega$ or $a$.

Therefore, *Sagnac effect can be explained considering that the mean light speed in a closed circuit is $c_0$ in the local atomic reference frame and that light speed is independent of the motion of the interferometer*. There is no experimental or observational result that contradicts these two statements and there is no other property of the speed of light supported by Sagnac effect. As the first one is already established, what we conclude now is that:

- *light velocity in empty space is independent of its source motion.*

We will name this the **property of light speed independence**. It means that in each point of the Universe, the (one-way) velocity of light in empty space can depend on the position of the point in time-space, on the local gravitational field, on the overall distribution of matter, on some unknown characteristic of the Universe, but does not depend on the motion of its source.

### 3.2. "at absolute rest"

Light speed being independent of source motion implies that one can consider an $R$ observer with a motion such that one-way light speed in relation to him is isotropic (in the absence of field, disregarding an eventual anisotropic positional dependence of light properties), whatever of light source. We will consider that this observer, as well as his reference frame, is "*at absolute rest*", by definition. We will identify such observer by the suffix 0, for instance, $R_0$. As we are considering a Euclidean space, because an $R$ observer has invariant measuring units and "instant vision", in any $R$ reference frame moving in relation to this one the measure of light speed is dependent on direction. The same cannot be stated in relation to an $A$ observer because this one measurements of one-way velocity depend on the method used for determining time coordinates. Einstein postulate of the constancy of light speed concerns an atomic observer and we will see later that there is no contradiction between Einstein postulate and the definition of absolute rest.

We will name velocity measures in $R_0$ as "*absolute velocities*". An $A$ observer with null absolute velocity (we are disregarding position dependence) has the same measuring units as $R_0$, as defined in §2.6. As the $A$ local mean light speed is $c_0$, so it is in $R_0$. Then, the one-way light speed in any reference frame at absolute rest is $c_0$. The one-way light speed in a reference frame at absolute rest is the absolute light speed. Therefore, by performing a local mean measure of light speed, an atomic observer obtains the value of the absolute light speed.

### 3.3. Angles and lines

The light speed source independence implies that the velocity of a light ray in relation to an observer depends on this one motion. The different angles and lines relevant for calculations are displayed in figure 2, being:

*Sight line*: defined by the vector $\boldsymbol{c_R} = \boldsymbol{c_0} \text{-} \boldsymbol{V}$.

*Light ray angles*: the angle $\phi$ between $\boldsymbol{V}$ and $\boldsymbol{c}$, being $\phi_0$ when relative to $\boldsymbol{c_0}$ and $\phi_R$ when relative to $\boldsymbol{c_R}$. Note that the difference between $\phi_0$ and $\phi_R$ is not a consequence of different measuring units but of the motion of the observer.

*Sight line angle*: the angle $\theta$ between $V$ and the sight line. It is the direction of the light source, when the observer is receiving a light ray, or of the emission ray, when the observer is the source.

*Connection line*: the line connecting two points of space in the same time moment in $R_0$. The angle between $V$ and the connection line is $\psi$.

*Position vector* **r**: the position vector of a point in relation to the observer.

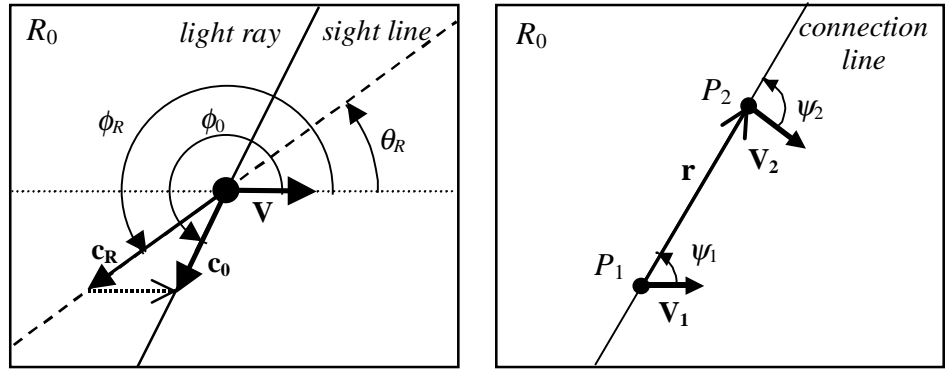

**Figure 2**. Definition of angles, lines and vectors.

### 3.4. Length and Time dependence with velocity

Consider a light ray covering a closed path in $A$ between two points, like in each arm of Michelson-Morley experiment. In $R_0$, the system is moving with absolute velocity and the light ray has the path represented in figure 3, being emitted at $P_1$ in the apparent direction of $P_2$, reflected in $P_2'$ and received in $P_1''$ From figure 3 one obtains:

$$l_0 = l_R \beta^2 \left( \gamma + V/c_0 \cdot \cos\phi_R \right) \tag{11}$$

where $\beta = \left( \sqrt{1 - V^2/c_0{}^2} \right)^{-1}$ and $\gamma = \sqrt{1 - V^2/c_0{}^2 \cdot \sin^2\theta_R}$. The time interval and the one-way light speed in the path $P_1 \rightarrow P_2$ is then:

$$t_R^{+r} = l_0/c_0 = l_R/c_0 \cdot \beta^2 \left( \gamma + V/c_0 \cdot \cos\phi_R \right)$$
$$c_R^{+r} = l_R/t_R^{+r} = c_0 \beta^{-2} \left( \gamma + V/c_0 \cdot \cos\phi_R \right)^{-1} \tag{12}$$

In the returning path, the expressions are the same, the angle being $\phi'_R$. Noting that $\phi'_R = \phi_R + \pi$, one obtains that the mean light speed in the path $P_1$-$P_2$-$P_1$ is:

$$\overline{c}_R = c_0 \beta^{-2} \gamma^{-1} \tag{13}$$

As, from the Michelson law, the mean light speed in $A$ is $\overline{c}_A = c_0$ and as $\overline{c}_A = \overline{c}_R /\left( LT^{-1} \right)$, one obtains:

$$LT^{-1} = \beta^{-2}\gamma^{-1} \tag{14}$$

Although we do not know what is the individual dependence with velocity of $L$ and $T$, we can parameterise such dependence. As Einstein, we will use a function $\varphi(V)$ such as:

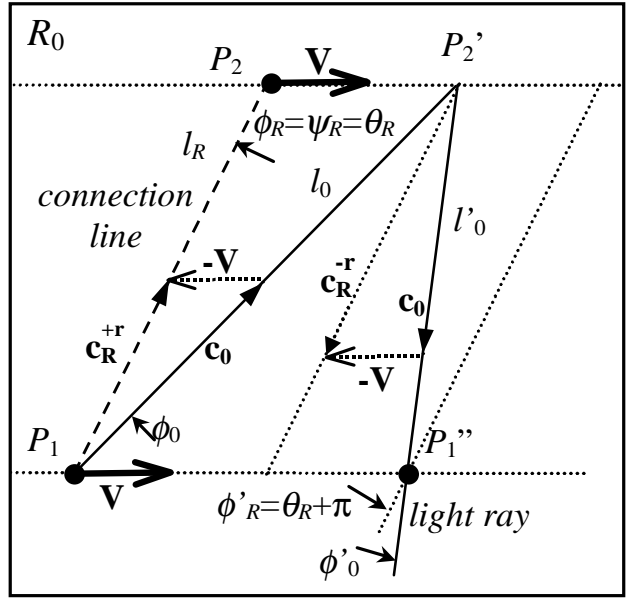

**Figure 3**. Representation in $R_0$ of a light ray that, in $A$, is emitted at $P_1$, reflected back in $P_2$ and received in $P_1$. The system $P_1$-$P_2$ as an absolute velocity $V$. The distance between $P_1$ and $P_2$ in $R$ is $l_R$. Note that the sight line corresponds to the connection line.

$$T = \varphi^{-1}\beta$$
$$L = \varphi^{-1}\beta^{-1}\gamma^{-1} \tag{15}$$

Some useful relations are:

$$\cos\theta_A = \gamma^{-1}\cos\theta_R$$
$$\sin\theta_A = \gamma^{-1}\beta^{-1}\sin\theta_R \tag{16}$$
$$\gamma\beta = \left(\sqrt{1 - V^2/c_0^2 \cdot \cos^2\theta_A}\right)^{-1}$$

Therefore

$$L = \varphi^{-1}\sqrt{1 - V^2/c_0^2 \cdot \cos^2\theta_A} \tag{17}$$

The one-way light speed is then:

$$c_A^r = \frac{c_0}{1 + V/c_0 \cdot \cos\phi_A} \tag{18}$$

which corresponds to the one-way light speed in the Zhang's analysis of Edwards [24] theory for $q_r = V/c_0 . \cos\theta_A$. The aberration laws are:

$$\cos\phi_0 = \frac{\gamma^{-1}\cos\phi_R + V/c_0}{1 + V/c_0 \cdot \gamma^{-1}\cos\phi_R} = \frac{\cos\phi_A + V/c_0}{1 + V/c_0 \cdot \cos\phi_A}$$
$$\cos\phi_A = \frac{\cos\phi_0 - V/c_0}{1 - V/c_0 \cdot \cos\phi_0} \tag{19}$$

The variation of $LT^{-1}$ (equation 14) implies a variation of the shape of bodies and, except for $\varphi = \beta$, a variation of the atomic time unit. In spite of these variations, space is still Euclidean. In LR theoretical framework, the procedure used by Einstein to determine the function $\varphi(V)$ is not valid because it requires the non-local validity of physical laws, what is not presumed in LR.

To explain the local invariance of phenomena in relation to an atomic observer, i.e., to satisfy LRP, the $L$ and $T$ dependence with velocity is not enough: other physical parameters have also to vary, according to LR conditions. To determine such variation is not important at this point.

### 3.5. Time coordinates

Consider an atomic observer with absolute velocity $V_1$ and a body with absolute velocity $V_2$. The atomic observer depends on light to know the spatial position of the body. Figure 4 represents the situation.

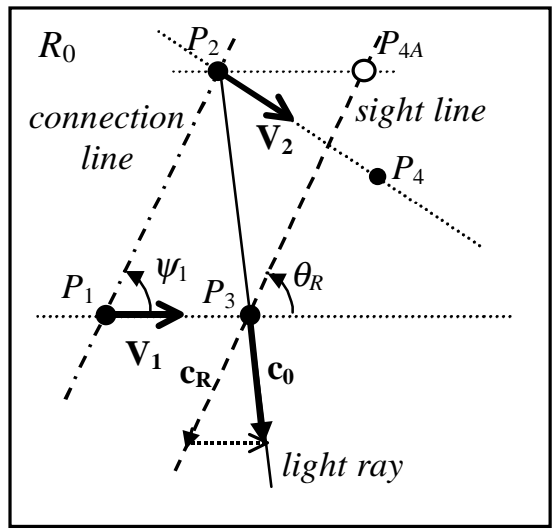

**Figure 4**. In the moment $t_0$, the atomic observer is at $P_1$ and the body at $P_2$. At that moment a light ray exits the body. In the moment $t'_0$, the light ray is received by the observer. At that moment, the observer is at $P_3$ and the body at $P_4$. The atomic observer "sees" then the body at $P_{4A}$ (open circle). As $\theta_R = \psi_1$, the $A$ observer "sees" the body in the relative position it had when the light ray exited it.

When the observer receives the light ray from the body, in the moment $t'_0$, he "sees" the body in its relative position at the moment the light ray exited the body, the moment $t_0$. In figure 4, the body is represented by a point but if one considers a body with a size, then one concludes that it is "seen" with the same angular size it had in relation to the observer at $t_0$, i.e., at the same distance.

To analyse the space position of moving bodies in relation to an atomic observer, we will consider a system of coordinates formed by the position vector **r,** with origin in the observer, characterised by its size or modulus and by the angle between observer's absolute velocity and position vector. The measure of $r$ modulus is not a problem (theoretically) because the observer can mark the space around him using a rigid measuring rod at rest in relation to him. In order to determine the correspondent time instants, i.e., the time coordinate, the observer has to calculate the time spent by light between the body and him, because what he knows is the time instant he received the light ray in his proper time units

The time spent by a light ray from a point with a position vector $r_A$ is, from (18) (note that $\phi_A = \theta_A + \pi$):

$$t_A^{-r} = \frac{r_A}{c_A^{-r}} = \frac{r_A}{c_0}\left(1 - V_1/c_0 \cdot \cos\theta_A\right) \tag{20}$$

or

$$t_A^{-r} = \frac{r_A}{c_0} - \frac{\mathbf{r_A} \cdot \mathbf{V_1}}{c_0{}^2} \tag{21}$$

where $(\mathbf{r_A} \cdot \mathbf{V_1}) = r_A V_1 \cos\theta_A$ is the internal product of the two vectors. When the $A$ observer knows his absolute velocity, then he can use (20); when he does not know that, he can presume that light speed in relation to him is $c_0$ or he can simply neglect the time spent by light. To each of these three cases there corresponds different values for time coordinates. In the first case we will name the time coordinates by Atomic time coordinates and represent them by $t_A$; in the second case, we name them Einstein time coordinates ($t_E$); and in the third case, Newton time coordinates ($t_N$). When referred to a specific point, this one is indicated by a superscript; when referring to the observer (proper time), the superscript is $O$. The relation between these time coordinates and proper time is:

$$\begin{aligned} t_A &= t_A^O - \frac{r_A}{c_0} + \frac{\mathbf{r_A} \cdot \mathbf{V_1}}{c_0{}^2} \\ t_E &= t_A^O - \frac{r_A}{c_0} \\ t_N &= t_A^O \end{aligned} \tag{22}$$

### 3.6. Measures of Velocity

Consider an observer $A$ with constant absolute velocity $V_1$ and a body $B$ with constant absolute velocity $V_2$. In $R_0$, at two successive instants $t_0$ and $t'_0$, the positions of body and observer define the vectors $r_R$ and $r'_R$. The relative velocity of the body in $R$ units is

$$\mathbf{v_R} = \mathbf{V_2} - \mathbf{V_1} = \frac{d\mathbf{r_R}}{dt_R} \tag{23}$$

Now, let us analyse the measures of velocity an $A$ observer can make. As we are considering three different time coordinates, to each one there corresponds a different time interval:

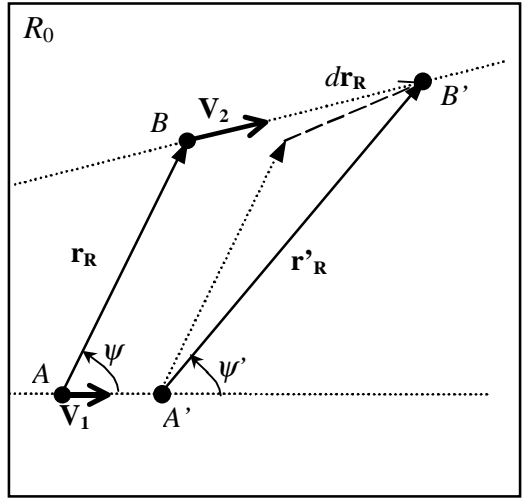

**Figure 5**. The positions in $R_0$ of an observer $A$ and a body $B$ in two time instants and their relative variation.

$$\delta t_A = \delta t_A^O - \frac{\delta r_A}{c_0} + \frac{\delta(\mathbf{r_A} \cdot \mathbf{V_I})}{c_0{}^2}$$

$$\delta t_E = \delta t_A^O - \frac{\delta r_A}{c_0} \qquad (24)$$

$$\delta t_N = \delta t_A^O$$

where $\delta$ is the scalar variation, for instance $\delta r = r' - r$ and $\delta t = t' - t$. The atomic velocity being

$$\mathbf{v_A} = \frac{d\mathbf{r_A}}{dt_A} \qquad (25)$$

the Einstein velocity is:

$$\mathbf{v_E} = \frac{d\mathbf{r_A}}{dt_E} = \frac{\mathbf{v_A}}{1 - (\mathbf{v_A} \cdot \mathbf{V_I})\big/ c_0{}^2} \qquad (26)$$

and the Newton velocity is:

$$\mathbf{v_N} = \frac{d\mathbf{r_A}}{dt_N} = \frac{\mathbf{v_E}}{1 + \dfrac{\mathbf{v_E}}{c_0} \cdot \dfrac{\mathbf{r}}{r}} \qquad (27)$$

Note that when the velocity to measure is the one-way light velocity, then the Einstein velocity is $c_0$, i.e.,

$$v_A = c_A^r \Rightarrow v_E = c_0 \qquad (28)$$

Therefore, *the Einstein one-way velocity of light is a constant with the value of local mean light speed*.

### 3.7. Collinear motion

When the motion of two bodies is such that their connection line has only a parallel transport, the motion is said to be collinear. In this case:

$$V_1 \sin\psi_{1R} = V_2 \sin\psi_{2R} \tag{29}$$

When one of the bodies is the observer, the motion of the other body is along the same sight line, i.e., $\theta = $ constant. Making

$$V_1^{r_A} = V_1 \cos\psi_{1A1}$$
$$V_2^{r_A} = V_2 \cos\psi_{2A2} \tag{30}$$

then the $A$ measures of velocity of the body are:

$$v_A^r = \frac{V_2^{r_A} - V_1^{r_A}}{1 - \left(V_1^{r_A}/c_0\right)^2}$$

$$v_E^r = \frac{V_2^{r_A} - V_1^{r_A}}{1 - V_2^{r_A} \cdot V_1^{r_A}/c_0^{\,2}} \tag{31}$$

$$v_N^r = \frac{V_2^{r_A} - V_1^{r_A}}{\left(1 + V_2^{r_A}/c_0\right)\left(1 - V_1^{r_A}/c_0\right)}$$

### 3.8. Doppler effect

The description of Doppler effect in $R$ is the classical one. The Doppler effect in $A$ is obtained from the Doppler effect in $R$ just by changing to the atomic measuring units. Consider an observer with absolute velocity $V$ and an electrodynamic wave from a distant source so that the angle $\phi$ between $V$ and the wave-normal is constant. The relative velocity of the wave along the wave normal is:

$$c_R^w = c_0 - V\cos\phi_0 \tag{32}$$

Therefore the relation between the frequency of the wave in $R_0$, $f_0$, and in $R$ is:

$$f_R = f_0\left(1 - V/c_0 \cdot \cos\phi_0\right) \tag{33}$$

In $A$, once the measuring unit of frequency is the inverse of time unit, it is:

$$f_A = f_0 \cdot \varphi^{-1}\beta\left(1 - V/c_0 \cdot \cos\phi_0\right) \tag{34}$$

For $\varphi = 1$ this is the special relativity equation. Considering the aberration laws (19) one obtains:

$$f_A = f_0 \cdot \varphi^{-1}\beta^{-1}\left(1 + V/c_0 \cdot \cos\phi_A\right)^{-1} \tag{35}$$

Considering now two atomic observers $A1$ and $A2$, with absolute velocities $V_1$ and $V_2$, from (35) one obtains the relation between the frequency measures of each observer:

$$f_{A1} = f_{A2} \cdot \frac{\varphi_2\beta_2\left(1 + V_2/c_0 \cdot \cos\phi_{2A2}\right)}{\varphi_1\beta_1\left(1 + V_1/c_0 \cdot \cos\phi_{1A1}\right)} \tag{36}$$

For a collinear motion, being $A2$ the wave source, using relation (16) for $\gamma\beta$ and noting that, in this case, it is $\gamma_1 = \gamma_2$, one obtains:

$$f_{A1} = f_{A2} \frac{\varphi_2}{\varphi_1} \sqrt{\frac{1 - v_E^r/c_0}{1 + v_E^r/c_0}} = f_{A2} \frac{\varphi_2}{\varphi_1} \sqrt{1 - 2 v_N^r/c_0} \tag{37}$$

where the velocities are relative to $A1$. In an experiment like Ives-Stillwell [25], where the wave emitted by $A2$ is received by $A1$ directly and also after reflection by a mirror in the opposite direction and at rest in relation to $A1$, the result obtained applying successively (37) is:

$$f'_{A1} = f_{A1} \cdot \frac{1 - v_E^r/c_0}{1 + v_E^r/c_0} \tag{38}$$

Doppler effect, dependent on absolute velocities, can therefore be represented by a simple physical law disregarding the absolute velocities. This is obtained by using measures of velocity that disregard the one-way velocity of light, considering whatever value for this one provided that it is the same in both ways. As the mean light speed is $c_0$, only Einstein measure of velocity can be compatible with consistent electrodynamic's laws independent of absolute velocities.

### 3.9 Special Relativity

The above property of Doppler effect is not an isolate case. For instance, the relation between sight line angles for two observers is easy to obtain in the particular case where the angle $\phi_0$ is the same for the two observers, which corresponds to $V_1//V_2$. In that case:

$$\cos \phi_{A2} = \frac{\cos \phi_{A1} - v_E/c_0}{1 - v_E/c_0 \cdot \cos \phi_{A1}} \tag{39}$$

Therefore, at least some phenomena can be described by physical laws independent of the absolute motion of the observer. This result is a consequence of the constancy of local mean light speed, or Michelson law. If one now considers a system of coordinates where time coordinates are obtained considering that one-way light speed is $c_0$, then $v_E$ in equations (38) and (39) gives place to the usual definition of velocity (the ratio between the variation of length and time coordinates). Therefore, some or all phenomena can be represented by physical laws independent of the inertial translational motion of the observer in a system of coordinates where the measure of light speed is constant *This conclusion corresponds to the two principles of Special Relativity*.

We have obtained equation (38) from the physics of Doppler effect, considering Michelson law. Because of that, we are aware of the particular nature of $v_E$. If, instead, we had only experimental results and ignored the physics of the phenomenon, we could have reached the same equation, however ignoring the particular nature of $v_E$. Analysing the law so obtained, we could then verify that it keeps invariant in a Lorentz coordinates transformation. This is what happens with Maxwell laws and Einstein analysis *of general laws of nature as they are obtained from experience*. Because the reference frame that supports Lorentz transformation is such that in it the measure of light velocity is always constant, the Einstein postulate of light speed constancy holds in it. Therefore, Michelson law and light speed source independence lead to light speed constancy in Einstein reference frame. However, the reason for such constancy is not the result of some connection between Time and Space.

From the above, we can now understand why light speed is constant in Maxwell laws: *it is the Einstein velocity of light*, which, as we have already noted, has the value of the local mean light speed, $c_0$. Equation (38) for Doppler effect was obtained in a 3-Euclidean space, with no connection between Time and Space. Therefore, the fact that equation (38) is invariant in a Lorentz transformation, as it happens with Maxwell laws, does not imply any particular characteristic for space-time.

Einstein Special Relativity, because it is based *in laws of nature obtained from experience*, namely Maxwell laws, does not reach the physical understanding that is possible in Local Relativity framework. Maxwell laws correspond to equation (38) for Doppler effect: they can fit observational results but are not a physical description of the phenomenon. Doppler effect is physically described as a consequence of the relative velocity of light and of the variation of observer's units with velocity. We have not yet reached a knowledge of electrodynamics that allow us to establish the physical description of the phenomenon. This suggest us that is not directly from Maxwell laws that one can expect to obtain a sound electrodynamics theory, in spite of the present success of electromagnetic theory.

An important aspect that this analysis evidences is that one cannot expect to develop a sound physical theory directly from *general laws of nature as they are obtained from experience*. Such a theory is just a first step in the analytical work because observational data is not absolute but relative to an atomic observer. Special Relativity and Ptolemy's theory are just two of the many present and past such theories. They can be useful but one must go beyond them.

## 4. Reference frames

There are reasons for considering several reference frames. One is that the appropriated calibration methods are not always possible or practical, as in length calibration for astronomical distances. Another one is that the observer lacks essential data for Time calibration, namely the knowledge of his absolute velocity. A different kind of reason is that an appropriate choice of the reference frame can greatly simplify the analysis. For instance, due to the relativity property exhibited by phenomena, an atomic reference frame can be more useful than an $R$ reference frame. Also, as we have seen in the analysis of Doppler effect, possibly because the variation of atomic properties with velocity is dependent on the mean light speed and not on the one-way light speed, an atomic reference frame with Einstein time coordinates can lead to much simpler analyses. Here we will analyse two atomic reference frames. The first one uses Atomic time coordinates, defined in the first of equations (22); we name it the $A$ reference frame. The other one has Einstein time coordinates (the second of equations (22)); we name it the Einstein reference frame. In both, length calibration is made by means of rigid measuring rods at rest in the reference frame.

### 4.1. Reference frame A

The coordinates of an $A$ reference frame correspond to the coordinates of the associated $R$ reference frame (i.e., the $R$ reference frame with axis coincident with the axis of the $A$ reference frame) modified by the relation between $A$ and $R$ measuring units. Note that there is no need to present any definition of synchronism for an $R$ observer because this observer is not dependent on light for acknowledging an occurrence: he has "instant vision".

Noting that $x_A = x_R / L^x$, $y_A = y_R / L^y$ and $t_A = t_R / T$, the coordinates transformation between $A$ and $R$ reference frames with coincident axis is:

$$x_A = \varphi \, \beta \, x_R$$
$$y_A = \varphi \, y_R$$
$$t_A = \varphi \beta^{-1} \, t_R \qquad (40)$$

The *ZZ* axis is identical to the *YY* axis and we will not refer to it. Although a spherical coordinates system may be the most appropriate, we use a system of orthogonal axis to ease the comparison with Special Relativity. Now, let us consider a second *A* reference frame and the associated *R* reference frame. At *t*=0 the axis of all reference frames are coincident and the relative motion is along *XX* axis. This is a simplifying assumption. We will use it because our objective is not to establish the general theory of atomic reference frames but to clarify the relation between Time and Space. Between the *R*1 and *R*2 reference frames, once measuring units are the same, only the *XX* coordinate is different, being:

$$x_{R2} = x_{R1} - (V_2 - V_1) \, t_R \qquad (41)$$

where $V_2$ and $V_1$ are the absolute velocities. Using (40) and (41) one obtains the coordinates transformation between two atomic reference frames that are coincident at *t*=0 and are moving along the *XX* axis:

$$x_{A2} = \frac{\varphi_2}{\varphi_1} \frac{\beta_2}{\beta_1} \left( x_{A1} - v_{A1} t_{A1} \right)$$
$$y_{A2} = \frac{\varphi_2}{\varphi_1} \, y_{A1} \qquad (42)$$
$$t_{A2} = \frac{\varphi_2}{\varphi_1} \frac{\beta_1}{\beta_2} t_{A1}$$

where $v_{A1} = \beta_1{}^2 (V_2 - V_1)$ is the measure of *A*2 velocity by *A*1 (ratio between the variation of length and time *A*1 coordinates of a point at rest in *A*2). It corresponds to the atomic collinear velocity, already established (31), being in this case $\psi_1 = \psi_2 = 0$.

The time coordinate transformation in (42) shows that if two occurrences have the same time coordinate in one absolute reference frame, the same happens in all absolute reference frames, i.e., simultaneity is absolute.

To determine the *A* reference frame we have yet to determine the function $\varphi(V)$. In the framework of Local Relativity one cannot state that the coordinates transformation has to be independent of absolute velocities and the value of $\varphi(V)$ cannot be determined as Einstein did in Special Relativity. The determination of $\varphi(V)$ is not necessary for the purpose of the paper and we will not analyse it here.

### 4.2. Einstein reference frame

We will now analyse the atomic reference frame where time coordinates are established considering that the one-way light speed is $c_0$. The importance of such a reference frame is the possibility that non-local physical phenomena may be described in it by physical laws independent of absolute velocity, as it happens with Doppler effect, therefore verifying Einstein Relativity Principle.

This kind of reference frame is sometimes referred as Lorentz reference frame because it supports Lorentz coordinates transformation, being the reference frame of Special Relativity.

We will name it *Einstein reference frame* and we define *an Einstein observer* as an atomic observer that determines the time coordinate of an occurrence at a distance $r_A$ considering that the time light spends between the occurrence and the observer is $r_A/c_0$. .

For an $R$ observer, who has "instant vision", two clocks are synchronised when he "sees" them marking the same. For an Einstein observer, two clocks are synchronised when he sees a difference between the clocks that is equal to the distance between clocks divided by $c_0$. The time light takes from a point at a distance $r_A$ to the origin is, from (20):

$$\Delta t_A = \frac{r_A}{c_A^{-r}} = \frac{r_A}{c_0}\left(1 - V/c_0 \cdot \cos\theta_A\right) \tag{43}$$

On synchronising the clock with the time difference of $r_A/c_0$, this Einstein observer introduces an error on the clock, as observed by $R$, of:

$$\Delta t_A - \frac{r_A}{c_0} = -\frac{V}{c_0^{2}}x_A \tag{44}$$

All clocks marking a difference, as seen by $R$, according with (43), are considered synchronised by the Einstein observer. For an $R$ observer, the $E$ (Einstein) clocks are not synchronised. We will name this difference (44) the *Einstein synchronisation bit*. Furthermore, as (44) is a linear function of $x$, the clocks are considered synchronised by any Einstein observer moving with the same velocity, fulfilling Einstein conditions for synchronism.

The $E$ time coordinate is the atomic time coordinate plus the synchronisation bit (44):

$$t_E = t_A - \frac{V}{c_0^{2}}x_A \tag{45}$$

Equation (45) is immediate from (22) but here we wanted to clarify the meaning of the synchronization bit.

Equation (45) shows that an Einstein reference frame exhibits dependence between time and space: its time coordinate is a function of length coordinate. It is also a function of velocity. As Einstein concluded, a velocity change affects clocks "synchronisation", i.e., clocks are considered to be "synchronised" by all Einstein observers relatively at rest but not by Einstein observers with different motions.

### 4.3. Lorentz transformation

As we have done for the transformation between $A$ reference frames, we begin by the coordinates transformation between $E$ and $R$ reference frames with coincident axis. Noting that $t_A = t_R/T$, from (40) and (45) it is:

$$x_E = \varphi\,\beta\,x_R$$
$$y_E = \varphi\,y_R \tag{46}$$
$$t_E = \varphi\beta^{-1}\left(t_R - \frac{V}{c_0^{2}}\beta^{2}x_R\right)$$

Note the correspondence with the change of variables presented by Lorentz in his article of 1904 [26]. The invariance of Maxwell laws with such a change of variables means that they have the same form either expressed in function of absolute unities as expressed in function of the coordinates of an Einstein reference frame. This implies that they are invariant whatever the Einstein reference frame. In the following, we will analyse, as before, the simple case of two observers with absolute velocities along their connection line.

Let us consider a second Einstein reference frame and the associated $R$ reference frame. Using (41) and (46) one obtains the coordinates transformation between two Einstein reference frames that are coincident at $t=0$ and are moving along the $XX$ axis with relative uniform Einstein velocity $v_E$:

$$x_{E2} = \frac{\varphi_2}{\varphi_1}\,\beta_E\left(x_{E1} - v_E t_{E1}\right)$$

$$y_{E2} = \frac{\varphi_2}{\varphi_1}\,y_{E1} \tag{47}$$

$$t_{E2} = \frac{\varphi_2}{\varphi_1}\,\beta_E\left(t_{E1} - v_E \frac{x_{E1}}{c_0{}^2}\right)$$

where $\beta_E \equiv \beta(v_E)$ and $v_E = \dfrac{V_2 - V_1}{1 - V_1 V_2 / c_0{}^2}$ is the measure of $E2$ velocity by $E1$ (ratio of length and time Einstein coordinates). It corresponds to the Einstein collinear velocity, already established (31), being in this case $\theta = 0$. A useful relation is $\beta_E v_E = \beta_1\beta_2(V_2-V_1)$.

To make the transformation (47) independent of absolute velocities requires to consider $\varphi(V) = constant$. As $\varphi(0) = 1$, by definition, this implies $\varphi(V) = 1$ (or, as Lorentz considered, that it "differs from unity no more than by a quantity of the second order"). Note that this does not correspond to any direct physical reality, neither does the constancy of the one-way light speed. We are just defining a practical reference frame, suitable for analyses of phenomena that can be described, in relation to each atomic observer, by equations that depend only on the mean light speed, like Doppler effect or electrodynamics. At this point, we do not know whether this is a general property of phenomena and if it applies not only to motion but also to field and position. Note also that (47) is valid only for a relative motion according with equation (41); as referred, this is a simplifying assumption, intended to support a simple analysis on the relation between time and space.

One remarkable characteristic of Lorentz transformation (47) is that (for $\varphi(V) = 1$), the transformation is independent of absolute velocities, in spite of the fact that Einstein reference frame (46) is dependent on absolute velocities.

### 4.4. Einstein reference frame in Local Relativity and in Special Relativity

Consider a light ray that exits the origin of an Einstein reference frame at $t = 0$, local time, and reaches a point $P$ at a distance $l$, where a local clock marks $l/c_0$. This is characteristic of an Einstein reference frame, no matter if the framework is Local Relativity or Special Relativity. However, in LR framework, where simultaneity is absolute, more affirmations can be produced. Consider that the Einstein reference frame was moving with absolute velocity $V$ along the $XX$ axis and that $P$ is on the positive $XX$ axis. In LR framework one can say that the clock at $P$ was marking $-lV/c_0{}^2$ when the light ray exited the origin and $l/c_0$ when the light ray reached it. At that instant, the clock in the origin was marking $(1+V/c_0).l/c_0$. One can therefore conclude that the light ray spent a time of $(1+V/c_0).l/c_0$ to cover the path, measured in the atomic time units of the observer, the relative velocity being then $\beta^2(c_0-V) = c_0/(1+V/c_0)$.

One must note that the only particularity of Einstein reference frame, besides implying a value for $\varphi(V)$ not yet determined in Local Relativity, is the synchronisation bit. The dependence of Length and Time with velocity is not a consequence of the constancy of the measure of one-way light speed but a consequence of the constancy of local mean light speed and of light speed independence in relation to source motion, having been obtained in LR framework.

*4.5. Time paradox in Lorentz transformation*

The characteristics of Einstein reference frame originate apparent time and length paradoxes that one can now easily understand, although some analytical work is required. The main conclusions are presented in italic.

First of all, one can note that for an $E$ observer the local clock of any other $E$ observer with a different velocity seems to run more slowly than his own. Consider two Einstein reference frames, $E1$ and $E2$, with absolute velocities $V_2>V_1$ along $XX$ axis. From (46), the relation between $E1$ and $E2$ time coordinates at $E1$ origin ($O1$) is:

$$t_{E2}^{O1} = \beta(v_E)\, t_{E1}^{O1} = \beta_E t_{E1} \tag{48}$$

To simplify the notation, when the coordinate is relative to the origin of the reference frame we will omit its indication. Equation (48) means that an observer at $E2$ concludes that time in $E1$ runs more slowly than in $E2$, at a constant rate $\beta_E$. Repeating the reasoning for $O2$, one obtains:

$$t_{E1}^{O2} = \beta(-v_E)\, t_{E2}^{O2} = \beta_E t_{E2} \tag{49}$$

Therefore, an observer at $E1$ concludes the opposite. This peculiar result is consequence of the velocity dependence of both the time unit and the synchronisation bit. From equation (45) that characterizes the Einstein time coordinate, noting that $t_{A2}=\beta_1\beta_2^{-1}t_{A1}$, one obtains:

$$t_{E1}^{O2} = \beta_2\beta_1^{-1}\, t_{E2} - \frac{V_1}{c_0^{\,2}} x_{E1}^{O2}$$
$$t_{E2}^{O1} = \beta_1\beta_2^{-1}\, t_{E1} - \frac{V_2}{c_0^{\,2}} x_{E2}^{O1} \tag{50}$$

In each of equations (50), the first term of second member represents the relation between proper times and the second term the synchronisation bit. For $E1$, $x_{E1}^{O2}>0$ and the synchronisation bit of $E1$ adds a retarding component that overcomes the shorter $E2$ time unit; for $E2$, $x_{E2}^{O1}<0$ and the synchronisation bit of $E2$ adds an accelerating component that is not enough to compensate for the larger time unit of $E1$. To show how the two effects, i.e., different proper times and synchronisation bit, compose to obtain the results (48) and (49), one has just to replace the length coordinate using the first of equations (47); for instance, for $E2$, noting that $x_{E1}=0$, it is

$$t_{E2}^{O1} = \beta_1\beta_2^{-1}\, t_{E1} + \frac{V_2}{c_0^{\,2}} \beta_E v_E\, t_{E1} = \beta_E t_{E1} \tag{51}$$

Therefore, *although the proper time of a moving clock in a Einstein reference frame can be slower or faster than the proper time of the observer*, depending on absolute velocities, *the synchronisation bit*, which transforms the proper time in the time coordinate, *is such that a moving clock is always delayed in relation to the clocks of the reference frame (time* *coordinates)*. The observer concludes that his proper time unit is shorter than the ones of moving bodies.

Now, if one considers two $E$ observers that execute a back and forth relative motion, one is induced from (48) and (49) to the conclusion that each one would see the other in retard when they return to initial positions, which is known as the *twin's paradox. The error* in this reasoning *is that Lorentz transformation* (47) *is valid only for a relative motion according to (41)*, where the origins of reference frames are coincident at $t$=0. Therefore, (47) is valid when the two observers get apart but not when they return. Note that this limitation on the validity of Lorentz transformation (47) is not a consequence of our particular deduction of the coordinates transformation.

To analyse this problem using Lorentz transformation, one can to modify (41) considering a generic initial position on $XX$ axis. The following generalised Lorentz transformation for $x$ and $t$ is obtained:

$$\Delta x_{E2} = \beta_E \left( x_{E1} - v_E \Delta t_{E1} \right)$$
$$\Delta t_{E2} = \beta_E \left( \Delta t_{E1} - v_E \frac{x_{E1}}{c_0^{\,2}} \right) \tag{52}$$

where $\Delta t$ and $\Delta x$ are the differences between the values of the coordinates of current point and of $E1$ origin at initial position. Using (52) one can now analyse the simple case where $E1$ is "at absolute rest" and $E2$ has a velocity $V$ at $t = 0$, when the origins are coincident, till the reversing point, when $E2$ inverts the motion, i.e., its velocity become $-V$. At the reversing point, which is the initial position for the returning motion, the $XX$ coordinate of $E1$ origin (in $E2$) is designated by $l_{E2}$ and proper times by $t_{E1,0}$ and $t_{E2,0}$.

Let us begin by analysing from $E1$. The relation between time coordinates at $E2$ origin ($x_{E2}$=0) is, whatever $t$:

$$t_{E1}^{O2} = \beta \, t_{E2} \tag{53}$$

This is an obvious result because $E1$ time coordinate has null synchronisation bit (null absolute velocity) and, as the absolute velocity of $E2$ has a constant modulus, his time unit is always $\beta$ times longer. Therefore, when the two observers meet each other, $E1$ sees the clock of $E2$ in retard:

$$t_{E1} = \beta \, t_{E2} \tag{54}$$

Now analysing from $E2$. The equations for time coordinates at $E1$ origin ($x_{E1}$=0) are:

$$t_{E2}^{O1} = \beta \, t_{E1} \qquad\qquad \text{for} \quad t_{E1} < t_{E1,0}$$
$$t_{E2}^{O1} = \beta \, t_{E1} + 2 \frac{V}{c_0^{\,2}} l_{E2} \qquad \text{for} \quad t_{E1} > t_{E1,0} \tag{55}$$

When the two observers meet each other ($x_{E2}^{O1} = x_{E1}^{O2} = 0$), $t_{E1} = 2 l_{E1}/V = -2 l_{E2}/(\beta V)$ and

$$t_{E2} = \beta^{-1} \, t_{E1} \tag{56}$$

Therefore, $E2$ concludes that the clock of $E1$ is in advance, in accordance with $E1$ conclusion (54). The use of generalised Lorentz transformation leads to a consistent result and time paradox disappears. However, although we have obtained consistency, we have not yet obtained understanding because one can still note that either when the two observers are moving away or when they are returning, for $E2$ it is always, from (52) or (55):

$$\partial t_{E2}^{O1} = \beta \; \partial t_{E1} \tag{57}$$

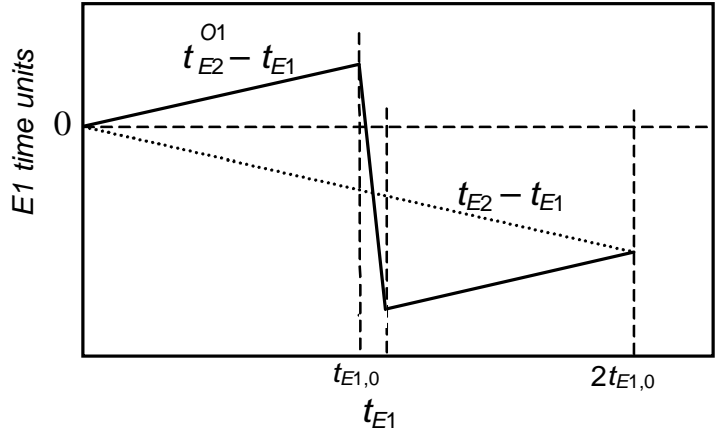

**Figure 6** – The difference between time coordinates of $E1$ origin. Axis in $E1$ time units. Dotted line represents the difference between $E2$ and $E1$ proper times. Solid line represents the difference between time coordinates of $O1$, which is the difference between proper times plus the synchronisation bit of $E2$. This synchronisation bit inverts signal when velocity changes, in a time interval of $l_{E1}/(c_0+V)$ in $E1$ time units. Except for this time interval, the varying synchronisation bit makes the smaller time unit of $E1$ to appear to $E2$ as being longer than his own.

The question is: if $E2$ always observes $E1$ clock running more slowly than its own, how does it happens that $E1$ clock is in advance when they meet? The answer is that when $E2$ inverts velocity, its synchronisation bit changes from positive to negative, retarding time coordinate, as equations (55) show (note that $l_{E2}<0$). This is evident using (50) (note that $x_{E2}^{O1} < 0$):

$$t_{E2}^{O1} = \beta^{-1} \, t_{E1} - \frac{V}{c_0^{\,2}} x_{E2}^{O1} \qquad \text{forward motion}$$

$$t_{E2}^{O1} = \beta^{-1} \, t_{E1} + \frac{V}{c_0^{\,2}} x_{E2}^{O1} \qquad \text{backward motion} \tag{58}$$

When his velocity changes, $E2$ will note that the clocks along $XX$ axis are no longer "synchronised" because *clocks "synchronisation" depends on velocity*, as noted before. *When $E2$ re-synchronises the clocks, he changes the time coordinate at $E1$ origin and $E1$ local clock becomes in advance*. This is a consequence of the fact that the time light takes between $E1$ and $E2$ changes because $E2$ is now moving toward the light ray and before it was moving away. It is also a consequence of the synchronisation method, of course, but *it is not a consequence of any physical phenomenon associated with the acceleration required for changing velocity*. One must note that between $t_{E1,0}$ and $t_{E1,0}+l_{E1}/(c_0+V)$ Lorentz transformation is not valid because the motion of the observers is not uniform, as required by equation (41). The velocity in the synchronisation bit is not $V$ or $-V$ but the mean value of $E2$ velocity in the time interval defined by light from $O1$ to $O2$, which, in this case, varies from $V$ to $-V$ in that time slot. The relation between $E2$ time coordinate of $E1$ origin and proper time of $E1$ is illustrated in figure 6.

### 4.6. Length paradox in Lorentz transformation

Using Lorentz transformation, the length of a rigid rod moving in an Einstein reference frame is always shorter than the length of an identical rod at rest in the reference frame and function of Einstein velocity. One can see from equations (45), which characterise Einstein reference frame, that the length of a rod depends only on the absolute velocity. The reason of the above result is that the measure of length is made considering the spatial positions of the extremes of the rod in the same time coordinate of Einstein reference frame and this is not the same instant for an $R$ observer. Length paradox is just a consequence of the synchronisation bit and the analysis of it is now trivial, so we will not present it here.

### 4.7. Attributing coordinates from observations.

An atomic observer acknowledges an occurrence by receiving information on it carried by a light ray or by a field. The information he obtains is relative to a time instant that is not the time instant of the reception. The time difference between the occurrence and reception moments depends on the relative distance between the occurrence and the observer and on the light speed relative to the observer. On observing a system of moving particles, the information received in one moment *does not portray the system in one moment but each particle in a different moment*. The practical consequences of this are very small and usually negligible except in two cases: in electrodynamics and in cosmic observations.

On the other hand, the information relative to a system received in the same point in time and space by two atomic observers with different motions is differently understood because they have different measuring units; if the observers are in difference points of space, also the time interval between an occurrence and its observation is different. Therefore, two different atomic observers present different descriptions of the same occurrence, and these descriptions cannot be transformed simply by a Galilean transformation. One consequence of this is that to interpret the interaction between fast moving particles, one has to obtain the description of phenomena in reference frames moving with each of the particles.

In what concerns cosmic observations from an earth observer, for a basic analysis one can neglect the consequences of observer motion and field. In this case, where the reference frame is "at absolute rest" (approximately) and field consequences are neglected, $R$, $A$ or Einstein reference frames of the earth observer are identical in LR framework. This is a convenient first scenario for the analysis of the positional dependence of physical parameters.

## 5. Conclusion (Part I)

The results already obtained can be summarised as follows.

Light speed is independent of source motion; it can depend on the overall matter distribution in the Universe, on the local gravitational field, on the time-space position, or/and on some unknown characteristic of the Universe, but not on its source motion. *This fundamental characteristic of light is stated in the property of light speed independence*. In this case, one shall expect atomic structure to depend on motion, implying atomic units dependent on motion. The problem is to characterise such dependence. Michelson-Morley experiment, together with other related experiments, can be explained considering that the geometry of atomic structure change with velocity in such a way that the time a light ray takes to go to a point at a fixed atomic distance and return is independent of direction and its measure in atomic time units seems to be constant. *This characteristic of atomic structure is stated in the Michelson law of the constancy of local mean light speed.* This suggests that atomic structure and atomic properties may depend on the relative bi-directional propagation speed of fields or of interaction particles. The change in atomic properties seems to be such that local physical laws, in relation to an atomic observer, are invariant. *This is stated in Local Relativity Property.* This property implies a relation between fundamental magnitudes and constants, expressed by *Local Relativity Conditions.*

The non-local analysis of physical systems with different motion, field or position is necessary to determine how do physical parameters vary in each case. To interpret non-local observations, the time-space structure has to be characterised.

The analysis of Doppler effect in LR framework shows that some (at least) non-local phenomena can be described by physical laws independent of the inertial translational motion of the observer in an atomic reference frame where time coordinates are established considering that one-way light speed is constant. *This is what the two postulates of Special Relativity state.* However, we have obtained this result in a 3-Euclidean space with no connection between space and time. Therefore, non-local observations can be interpreted using such structure for space and time. The Special Relativity line element is not characteristic of time-space but only of Special Relativity reference frame.

A number of problems are solved, like the interpretation of Sagnac effect, the Time-Space connection, the simultaneity concept and the mathematical and physical reasons of Time and Length paradoxes. It is important to note that the analysis here presented at no point conflicts (formally) with Einstein's article of 1905.